\documentclass[ 12pt]{iopart}
\usepackage{epsfig}
\def\be{\begin{equation}}
\def\ee{\end{equation}}
\def\ba{\begin{eqnarray}}
\def\ea{\end{eqnarray}}
\def\bL{\bar L}
\def\bC {\bar C}
\def\v{v_{dc}}

\begin{document}

\title{Improving high-T$_c$ dc-SQUID performance by junction asymmetry}
\author{Urbasi Sinha $^1$, Aninda Sinha $^2$ and Frank K. Wilhelm$^1$}
\address{$^1$ Institute for Quantum Computing and Department of Physics and Astronomy, University of Waterloo, \\ 200 University Avenue West, Waterloo, Ontario N2L 3G1,Canada}
\address{$^2$ Perimeter Institute for Theoretical Physics, \\  31 Caroline Street North, Waterloo, Ontario N2L 2Y5, Canada}
\ead{$^1$ usinha@iqc.ca}
\ead{$^2$ asinha@perimeterinstitute.ca}
\ead{$^3$ fwilhelm@iqc.ca}

\begin{abstract}
We study noise and noise energy of a high-T$_c$ dc SQUID fabricated on a high-$\epsilon_R$ substrate whose conduction properties are given by transmission line physics. We show that transmission line resonances greatly enhance the noise. Remarkably, resistance asymmetry enhances these resonances even more.
However, as the transfer function ($\frac{dv_{dc}}{d\phi}$) scales the same way, the noise energy is reduced by asymmetry greatly enhancing the flexibility and performance of the
SQUID.
\end{abstract}

\section{Introduction}

SQUIDs are versatile high-resolution magnetometers that find wide application in fields ranging from brain research, low-field MRI, geological prospecting, precision electrometry, detection of elementary particles and quantum computing. They are so powerful because of their large sensitivity and low intrinsic noise, leading to excellent signal-to-noise ratios that in some cases approach the limit set by quantum mechanics. High $T_C$ SQUIDs offer the possibility to be operated at temperatures of liquid nitrogen, lowering cost and enhancing flexibility. Their electrodynamics is more involved than that of their low-$T_c$ counterparts and optimization of signal-to-noise figures requires renewed attention.

 In a previous paper \cite{Urbasi1}, we have analytically studied transmission line resonances in high $T_C$ dc SQUIDS. Such resonances are exhibited in the characteristics of SQUIDs fabricated on substrates with high dielectric constant like strontium titanate. In \cite{Urbasi1}, we analytically derived the SQUID power balance equation for both symmetric and asymmetric SQUIDs and investigated SQUID current-voltage $I (V)$, voltage-flux $V (\Phi)$ and voltage modulation $\Delta V$ characteristics. In this paper, we analytically study the effect of transmission line inductance on the noise characteristics of a dc SQUID. We will closely follow the methods used in \cite{Enpukunoise}.

The paper is organized as follows. In section 2, we set notations by describing the relevant circuit equations. In section 3, we study the effect of asymmetry on white noise in transmission line dc SQUIDs. We conclude in section 4. The calculational details are provided in appendix A. The SQUID parameters are the same as those used in \cite{Urbasi1}.

\section{Circuit equations}
 The geometry of the SQUID washer that we have used is shown in fig.(1a) \cite{Urbasi1}. This is a geometry commonly used to manufacture SQUIDs \cite{Lee, Ludwig, Bar} and is the geometry Enpuku et al used to numerically investigate the effects of large dielectric constant of strontium titanate (STO) on the  characteristics of high $T_C$ dc SQUIDs \cite{enpuku}. The slit of the SQUID washer makes up the SQUID inductance where $l$, $s$, $w$ and $d$ denote the slit length, slit width, electrode width and thickness of electrode respectively. For this geometry, the inductance per unit length of the slit $\bL$ and parasitic capacitance per unit length $\bC$ are given by \cite{enpuku, Yoshida}:
\be
\bL=\bL_{M}+\bL_{K}\,
\ee
where $\bL_M$ is the magnetic inductance per unit length and $\bL_K$ is the kinetic inductance per unit length of the SQUID slit given by\cite{Ramo}:
\be
\bL_{M} ={ \mu_0 K(k)\over K(k')}\,
\ee
\be
\bL_{K} = {2 \mu_0 \lambda^2 \over d w k'^2 K^2 (k')} [{w\over s} \ln [{4 w s \over d (w+s)}] + {w \over 2w + s}\ln [{4 w (2w+s)\over d (w+s)}]]
\ee
and
\be
\bC ={ \epsilon_{R} +1\over 2c^{2}\bL_{M}}
\ee
where $\mu_0$ is the permeability of free space, $K(k)$ is the complete elliptic integral of the first kind \cite{math} with a modulus $ k=\displaystyle {s\over s+2w}, k' = (1-k^2)^{1/2}, \lambda$ is the penetration depth of the film, $\epsilon_R$  is the dielectric constant of the STO substrate and $c$ is the velocity of light in vacuum.
The SQUID slit behaves as a transmission line with distributed inductance $L$ and distributed capacitance $C$. The impedance $Z_{AB}$ of the slit seen from terminals A and B is given by \cite{enpuku}:
\be\label{impedance}
Z_{AB}=iZ_{0}\tan (\Omega l \sqrt{\bL \bC}) + i\Omega L_{PR}\,
\ee
where $Z_0=\sqrt{\bL/\bC}$ is the characteristic impedance of the slit transmission line, $\Omega$ is the angular frequency of measurement and $L_{PR}$ is the junction parasitic inductance. The first term arises because the hairpin shaped slit can be treated as a shorted transmission line of length $l$. \\
Now, using the formula \cite{enpuku}
\be
\tan ({\pi x\over 2})=-{4x \over \pi} \displaystyle\sum_{n=1}^\infty {1\over x^2-(2n-1)^2}\,,
\ee
eqn.(\ref{impedance}) can be expanded as (in the lossless case with $L_{PR} = 0$):
\be\label{ZAB}
Z_{AB} = \displaystyle\sum_{n=1}^\infty {1\over i\Omega C_{n} + {1/ i\Omega L_{n}}}\,,
\ee
with
$C_n = {\bC l/ 2}$ and $L_n = {8\bL l/ \pi^2 (2n-1)^2}$. This transformation to an equivalent circuit has the advantage that it allows us to consider loss in the transmission line. In the lossy case, the rf-loss $R_n$ is added to eqn.(\ref{ZAB}) leading to:\\
\be\label{lossyZAB}
Z_{AB} = \displaystyle\sum_{n=1}^\infty {1\over i\Omega C_n + {1/ i \Omega L_n} + {1/ R_n}}\,,
\ee
where
$R_n = Q \sqrt{{L_n/ C_n}}$ and $Q$ is a quality factor. Here $l$ is the SQUID slit length. This expression allows us to express the impedance $Z_{AB}$ by the series of L-C-R resonant circuits as shown in fig.(1b).
 The circuit equations can be easily derived by the application of Kirchhoff's laws as in \cite{Urbasi1}. In fig.(1b), current entering point C should equal current leaving point C. We are assuming the most general case in which the SQUID is made up of junctions which are asymmetric \cite{Koelle, Ed}. Here $J$ is the circulating current through the SQUID inductance, $I_B$ is the SQUID bias current, $V_1$ and $V_2$ are voltages across junctions 1 and 2 , $\theta_2$ and $\theta_1$ are the phases of junctions 2 and 1 and $R_D$ is a damping resistance in parallel to the SQUID inductance.

\begin{figure}[ht]\label{geometry}
\includegraphics[width=.5\textwidth, angle=-90]{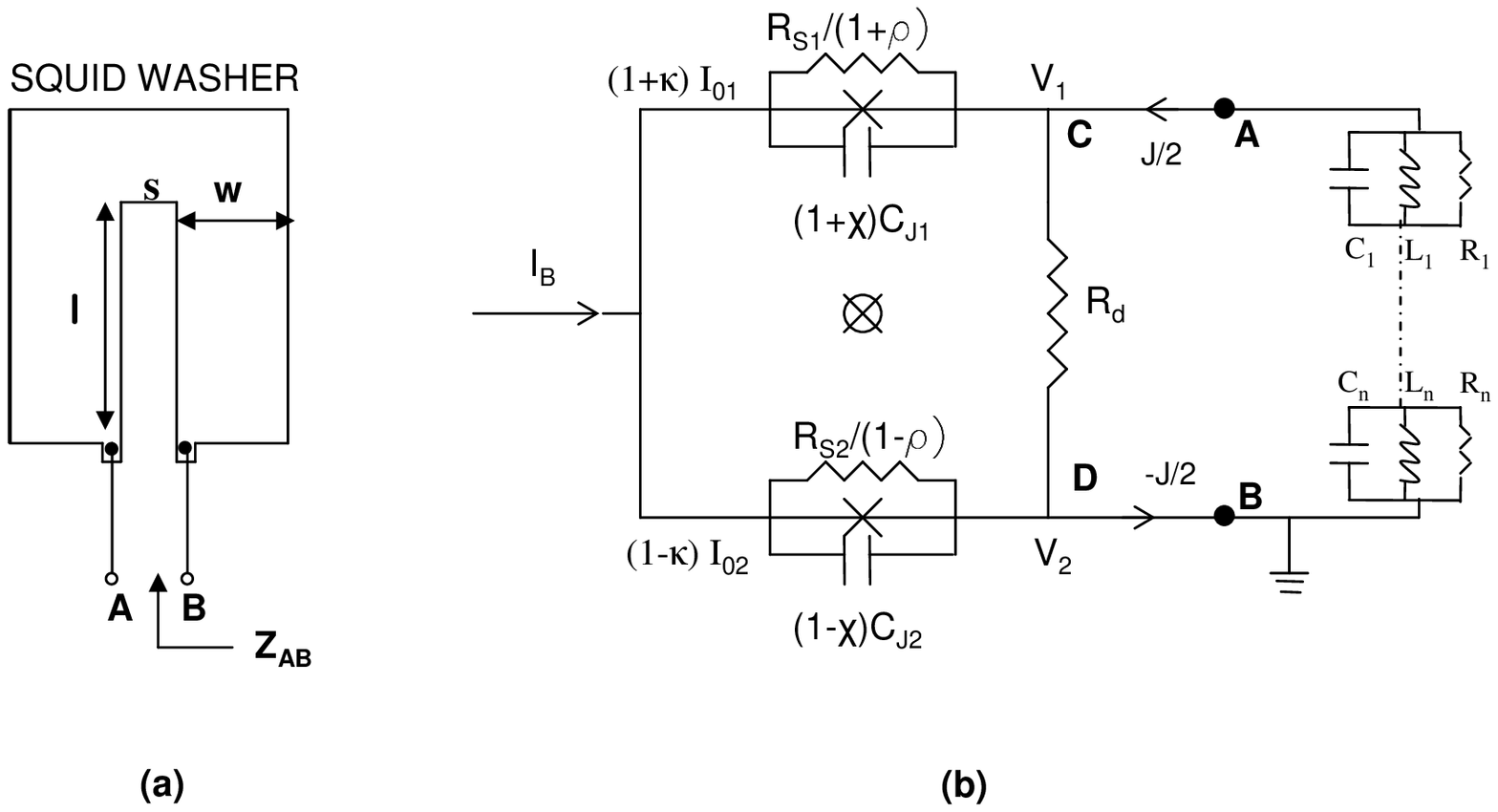}
\caption{ (a) Geometry of the SQUID washer. Here {\it l}  is the length of the slit, {\it w} is the width of the electrode and {\it s} is the width of the slit.  (b) Equivalent circuit of the washer when its parasitic capacitance distributing along the slit of the washer is taken into account. The circuit consists of the SQUID coupled to a series of LCR circuits. $R_d$ is a damping resistance in parallel to the SQUID. Here current asymmetry is denoted by $\kappa $, normal state resistance asymmetry by $\rho$ and capacitance asymmetry by $\chi$.}
\end{figure}

 We write down the normalized circuit equations for the SQUID loop including random noise currents $i_{n1}$ and $i_{n2}$. Let the average junction critical current be $I_0$, the average junction normal state resistance be $R_S$ and the average junction capacitance be $C_J$. Let the asymmetry parameter in $I_0$ be $\kappa$, that in $R_S$ be $\rho$ and that in $C_J$ be $\chi$. So specifically, let us split \cite{Koelle} $2 I_0=(1+\kappa)I_0+(1-\kappa)I_0, {2/ R_S}={(1+\rho)/ R_S}+{(1-\rho)/R_S}$ and $2 C_J=(1+\chi)C_J+(1-\chi)C_J$. We normalize currents by $I_0$, voltage by $I_0 R_S$ and time $t$ by ${\Phi_0/ 2\pi I_0 R_S}$. The ac Josephson relation gives  ${\it v_{dc}} = {d\theta / d \tau} = {V/I_0 R_S}$ , where $v_{dc}$ is the normalized voltage and ${\it \tau}$ is the normalized time. Then including random noise currents $i_{n1}$ and $i_{n2}$, the normalized circuit equations are:
\be\label{normalised1}
(1+\chi)\beta_C \ddot\theta_{1} = {1\over 2} (i_B + j) - (1+\rho)\dot\theta_{1} - (1+\kappa)\sin (\theta_1) - \gamma (\dot\theta_{1} - \dot\theta_{2}) + i_{n1}\,,
\ee
\be\label{normalised2}
(1-\chi)\beta_C \ddot\theta_{2} = {1\over 2} (i_B - j) -(1-\rho) \dot\theta_{2} - (1-\kappa)\sin (\theta_2)  +\gamma (\dot\theta_{1} - \dot\theta_{2}) + i_{n2}\,,
\ee
Here,   $\beta_C = {2 \pi I_0 C_J R_{S}^2/ \Phi_0}$ is the SQUID McCumber parameter with $\Phi_0$ being the flux quantum and $ \gamma = {R_S/ R_D}$. In \cite{Urbasi1} the current-voltage characteristics in the absence of noise was derived. It was found to be
\begin{eqnarray} \label{exact}
i_b&=&2\v + {I^2 \v \over 2 |d s-r \bar r|^2} \bigg\{ (1+\kappa^2) [d \bar d+r \bar r] +2 \kappa [d \bar r +\bar d r]\nonumber \\  &+&(1-\kappa^2)\bigg( [d \bar d-r \bar r]\cos (2\pi\phi)
   + i [d \bar r-\bar d r ]\sin(2\pi\phi)\bigg)\nonumber \bigg\} \\
   &+&  {I^2 \v \over 2 \beta^2 |d s - r \bar r|^2} (1+2\gamma- i{A-\bar A\over 2 \v})   \bigg\{ (1+\kappa^2) [\beta^2 s \bar s + r \bar r] - 2\kappa\beta [\bar s r + s \bar r]\nonumber \\ &-&(1-\kappa^2)\bigg ([\beta^2 s \bar s - r \bar r]\cos(2\pi\phi) - i\beta [s \bar r - \bar s r ]\sin (2\pi\phi)\bigg )\bigg\}\,.
\end{eqnarray}
Here $d = d(v_{dc}), s = s(v_{dc})$, $ r = r(v_{dc})$, $A=A(\v)$ and $\bar x$ denotes the complex conjugate of $x$. The phase difference between the two junctions have been set to $2\pi\phi$ where $\phi$ is the externally applied flux normalized to $\Phi_{0}$.
Here,
\be
I = \sqrt { 2 v_{dc} ( 1 + v_{dc}^2 )^{1/2} - v_{dc} }\,
\ee
\be\label{sv}
s(\omega) = ( \omega^2 \beta_C + i \omega )\,, \quad d(\omega)= A(\omega)-(\omega^2 \beta_C + i \omega + 2 i \omega \gamma)\,,
\ee
\be
r(\omega)=i\rho\omega+\chi\beta_C\omega^2\,,
\ee
and
\be \label{Anoloss}
A (\omega) ={2 \over \bL l } {\Phi_0 \over 2 \pi I_0} \frac{ ({\omega l 2 \pi I_0 R_S \sqrt{ \bL \bC}/\Phi_0})}{ \tan ({\omega l 2 \pi I_0 R_S \sqrt {\bL \bC}/\Phi_0})}\,.
\ee
If $L_{PR}\neq 0$ then $A(\omega)$ should be replaced by ${A(\omega)/ (1+{\pi I_0 A(\omega) L_{PR} / \Phi_0})}$ in the calculations \cite{Urbasi1}. $\bar L$ and $\bar C$ are the SQUID inductance per unit length and SQUID parasitic capacitance per unit length respectively. The dielectric constant $\epsilon_R$ enters through $\bar C$. For more details the reader is referred to \cite{Urbasi1, enpuku}.

\section{White noise}
We will now proceed to analyse the effect of noise with white power spectra, following closely the analysis in \cite{Enpukunoise}. The calculation details are given in the appendix. The algebra is straightforward but very tedious.
 The final formula can be expressed as
\be
s_v(\Omega)=s_{\delta i_b} r_d^2 (1+\alpha_1) +s_{\delta\phi}v_\phi^2 (1+\alpha_2)\,.
\ee
Here $r_d=\partial \v/\partial i_b$ and $v_\phi=\partial \v/\partial \phi$ and can be extracted from (\ref{exact}).  The noise power per unit angular frequency are given by
\begin{eqnarray}
s_{\delta i_b}(\omega)&=&{2\Gamma\over \pi}\,,\\
s_{\delta \phi}(\omega) &=& {\beta^2 \Gamma \over 8\pi}\,,
\end{eqnarray}
with $\Gamma=2\pi k_B T/I_0 \Phi_0$ a noise parameter, $k_B$ is the Boltzmann constant and $T$ is the temperature. As in \cite{Enpukunoise} we define the noise power spectra in practical units as
\begin{equation}
S_v=2 k_B T R_s\left[ 4 r_d^2(1+\alpha_1)+{\beta^2\over 4} v_\phi^2 (1+2\gamma)(1+\alpha_2)\right]\,,
\end{equation}
whose units are $V^2/Hz$. In the lumped case limit, the expressions in \cite{Enpukunoise} are reproduced.
The explicit formula for $\alpha_1$ and $\alpha_2$ are rather unwieldy  \cite{notebook}. Since we are interested in small asymmetry, we only quote the expressions to leading order in $\rho,\kappa,\chi$ with $\gamma=0$.
\begin{eqnarray}
\alpha_1&=&{\cos^2\pi\phi\over 8 r_d^2 d \bar d}+{\sin 2\pi \phi [(\kappa-\rho)\v^3 \beta_c-\rho A(\v) \v]\over 8 r_d^2 d \bar d s \bar s}\,,\\
\alpha_2 &=& {2 \sin^2 \pi\phi\over v_\phi^2 \beta^2 d \bar d}+ {4\sin\pi \phi[-(\kappa-\rho)\v^3 \beta_c \cos\pi\phi +\kappa s \bar s \sin\pi\phi ]\over v_\phi^2 \beta^2 d \bar d s \bar s }\,,
\end{eqnarray}
where $s, d$ are given in equations (\ref{sv}).
We note that the capacitance asymmetry does not appear explicitly in the above formula although it is implicitly present in $r_d$ and $v_\phi$.
In figure 2(a), we show plots of SQUID voltage noise vs. bias voltage for different values of dielectric constant $\epsilon_R$ for both when the SQUID inductance is taken to be a lumped element and when it is taken to be a transmission line. The SQUID parameters are the same as those used in \cite{Urbasi1} i.e. $l=55\mu m$, SQUID inductance, $L_{SQ}(\bar L \times l) = 55pH, I_0 = 4.75 \mu A, R_S = 13.8\Omega$, SQUID parasitic inductance $L_{PR}= 13pH$, $\beta_{C} = 0$ and damping resistance $R_D =0$.
Figure 2(a) shows that in the lumped inductance limit, the noise curves overlap with each other. Thus the noise as a function of bias voltage is independent of substrate dielectric constant in this case. However in references \cite{Urbasi1,enpuku,Lee}, it has been shown that a high substrate dielectric constant can cause transmission line resonances in dc SQUID characteristics. Thus, in case of high $T_C$ SQUIDs, which are usually fabricated on STO substrates that are known to have a very high dielectric constant, it is important to model the SQUID inductance as a transmission line. In this case, we can see in figure 2(b), that the SQUID noise is definitely a function of dielectric constant and a high dielectric constant causes resonances to appear in the SQUID voltage noise vs. bias voltage curves at low voltages. This can be understood as follows. $\sqrt{S_v}$ is maximum when $r_d$ is maximum. It follows from the analysis in \cite{Urbasi1} that this happens approximately at $\v=\displaystyle {(n-1/4) \Phi_0 \over 2 l I_0 R_S \sqrt{\bar L \bar C}}$ for positive integral $n$. Since $\bar C$ is inversely related to $\epsilon_R$, it follows that for lower values for $\epsilon_R$ the first extremum in the noise occurs at higher $\v$. 
Resonances start appearing at voltage $\v$ whenever the associated Josephson frequency matches the frequency of the lowest mode of the finite-length transmission line.

\begin{figure}[t]
\begin{tabular}{ll}
\hskip 2cm \includegraphics[width=.45 \textwidth]{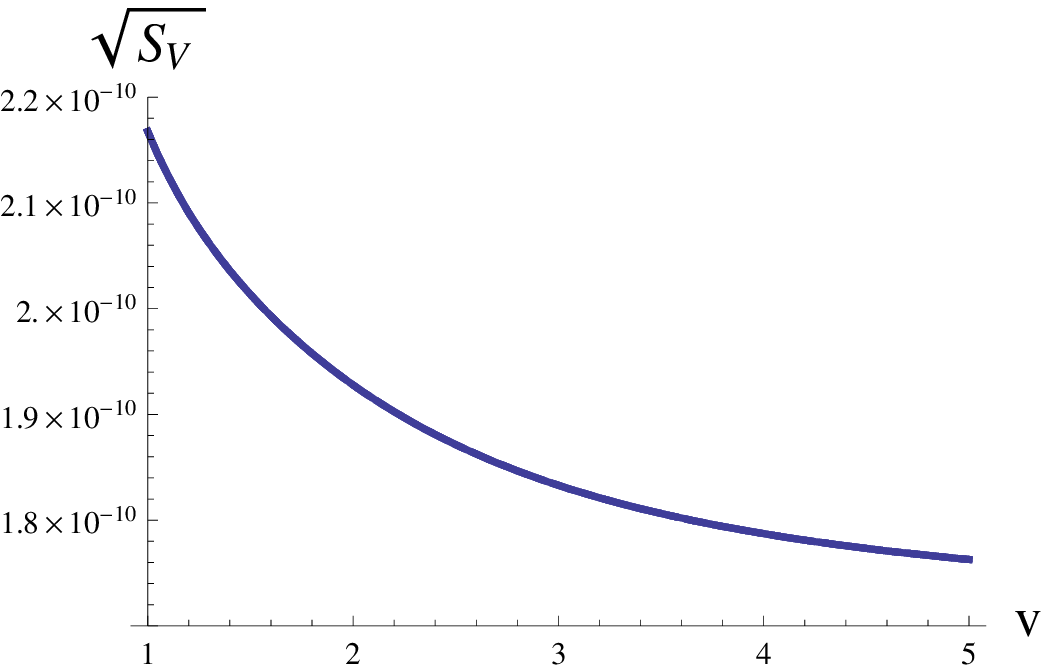}
&\includegraphics[width=0.45 \textwidth]{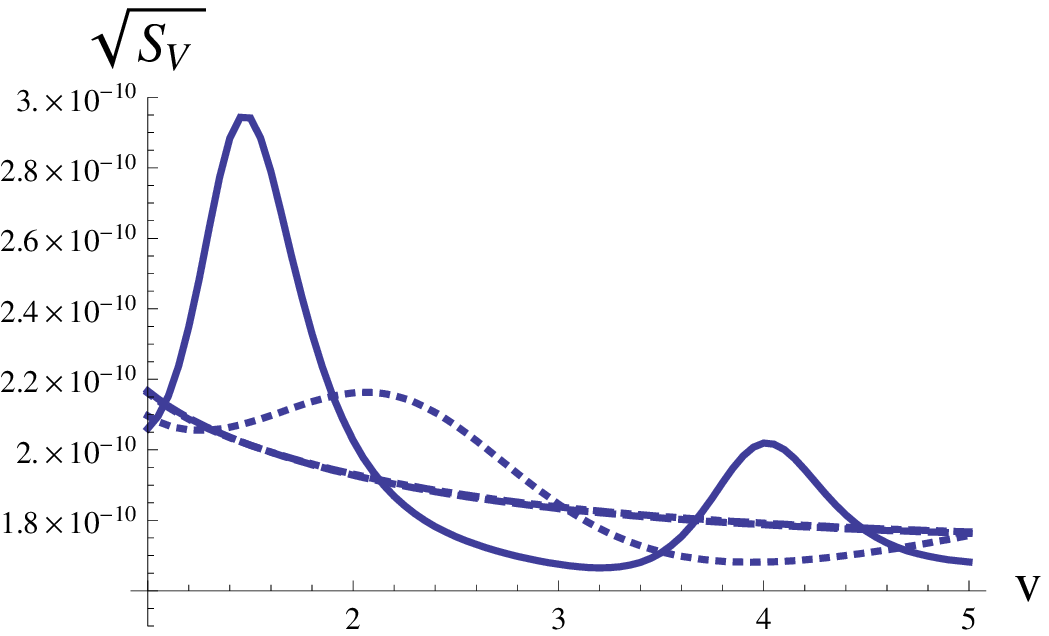}\\
\hskip 5cm (a)& \hskip 3cm (b)
\end{tabular}
\caption{(a) SQUID voltage noise vs. bias voltage  for different values of dielectric constant $\epsilon_R$ for the case when the SQUID inductance is taken to be a lumped element and not a transmission line. All the cases overlap which indicates that the noise value as a function of bias voltage is independent of $\epsilon_R$ in this case. (b) Graphs for different values of dielectric constant $\epsilon_R$ for the case when the SQUID inductance is taken to be a transmission line. The dashed line is for $\epsilon_R = 10$ and the dot-dashed line is for $\epsilon_R = 100$ and they overlap. The dotted line is for $\epsilon_R = 1000$ and the solid line is for $\epsilon_R = 2000$.}
\end{figure}

In figure 3, we show plots of SQUID voltage noise vs. bias voltage as a function of asymmetries in junction parameters. Here $\epsilon_R = 2000$ is used. Again we show both the lumped inductance limit as well the transmission line limit. We can see that in case of a lumped SQUID inductance, asymmetries in junction parameters do not affect the plots too much and one can say that presence of asymmetries leads to a marginal increase in noise as a function of bias voltage. However, when we consider the transmission line limit, asymmetries have a significant effect on the curves especially at resonance positions. $\rho$ asymmetry causes the sharpest increase. In our previous paper \cite{Urbasi1}, we have seen that this is also the case for the $dV/d\phi$ vs. bias voltage curves. Thus $\rho$ asymmetry enhances the peak in both $dV/d\phi$ as well as voltage noise curves.

\begin{figure}[ht]
\begin{tabular}{ll}
\hskip 2cm \includegraphics[width=.45 \textwidth]{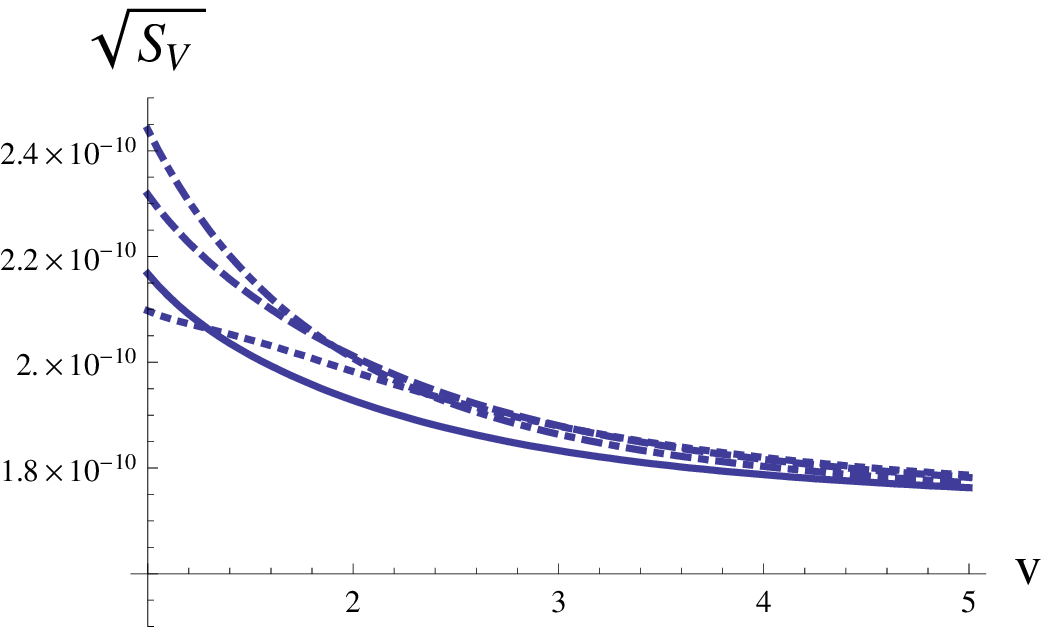}
&\includegraphics[width=0.45 \textwidth]{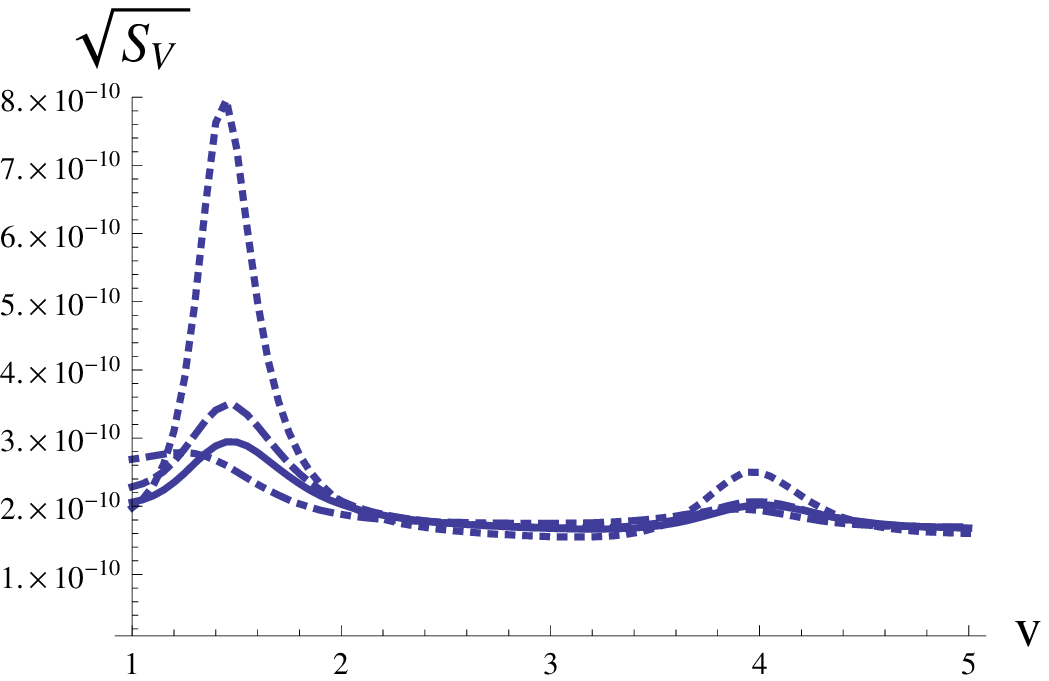}\\
\hskip 5cm (a)& \hskip 3cm (b)
\end{tabular}
\caption{(a) SQUID voltage noise vs. bias voltage for different asymmetry parameters and a fixed $\epsilon_R = 2000$ in the case of a lumped SQUID inductance. The solid line is for $\rho = \kappa = \chi =0$, the dotted line is for $\rho = 0.2, \kappa = \chi =0$, the dashed line is for $\kappa = 0.5, \rho = \chi =0$ and the dot-dashed line is for $\kappa = 0.5, \rho = 0.2$ and $\chi = 0$ (b) Graphs for the transmission line case where the different types of lines stand for the same combination of asymmetries as in the lumped case.}
\end{figure}

This can be understood as follows: As we have explained in the previous section, the resistance asymmetry $\rho$ has been chosen such that the total resistance of the SQUID would be a constant if the SQUID inductance behaved as a lumped element as opposed to a transmission line. At positions off-resonance, the conductance through the transmission line is aided by the finite width of the resonant peaks.  Thus off resonance, the same physics applies and the total off-resonant conductance of the SQUID remains constant. From the point of view of transmission line physics, the off-resonant conductance is proportional to the sum of the peak widths of the arms of the SQUID. On resonance, however, the conductance is proportional to the sum of ``quality factors", proportional to the {\em inverse} peak widths. This quantity increases as the asymmetry is increased. It needs to be emphasized that this is the asymmetry of the {\em junctions}, i.e., the {\em loads} of the transmission line and not asymmetries in the transmission lines themselves.

The effect is strongest for resistance asymmetry because it enters inversely into the quality factor. Critical current asymmetry enters through the Josephson inductance ($L_J \propto \frac{1}{I_0}$). It is known that quality factor $Q \propto R \sqrt {C/L}$. Thus critical current asymmetry enters under the square root and thus has a much smaller influence. Also, both effects due to critical current asymmetry and resistance asymmetry counteract themselves.

Let us now briefly discuss the noise energy defined as \cite{Enpukunoise,squidhandbook}
\begin{equation}
E=S_v(f)/2 L v_\phi^2 \left({I_0 R_s\over \Phi_0}\right)^2\,.
\end{equation}
Remarkably, even though this asymmetry increases the absolute noise level, the noise energy is generally lowered by resistance asymmetry. This is seen in figure 4 a and b. As introduced above, the noise energy is the appropriately normalised performance
quantifier, relating the absolute noise to the squared transfer function. As the transfer function shows the same enhancement by asymmetry just discussed but enters quadratically into the noise energy, the resistance asymmetry reduces the noise energy close to resonance and thus is rather smooth across the voltages of interest.

\begin{figure}[ht]
\begin{tabular}{ll}
\hskip 2cm \includegraphics[width=.45 \textwidth]{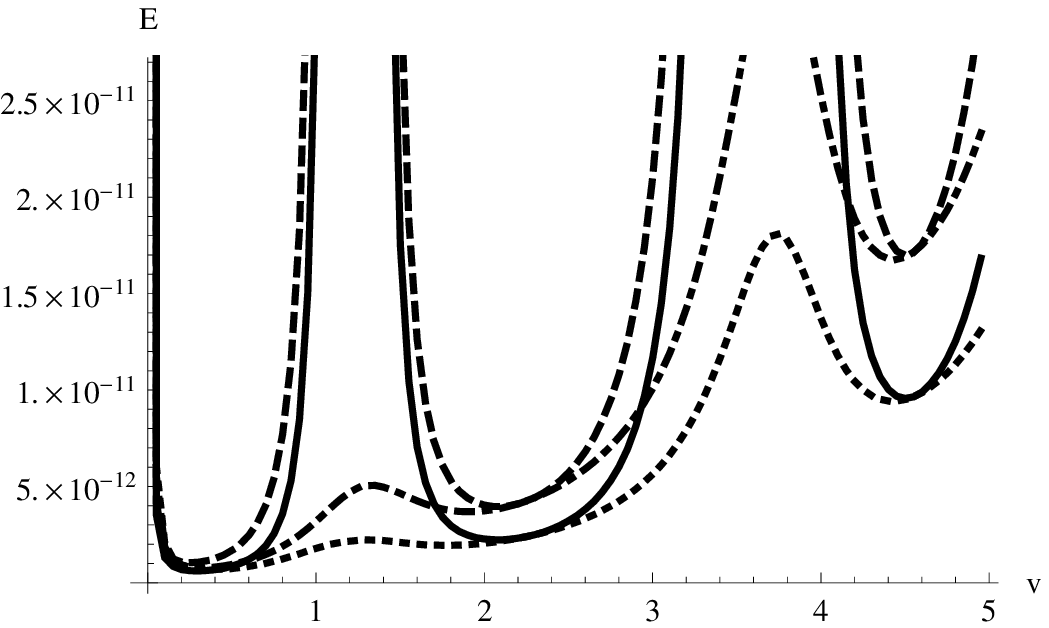}
&\includegraphics[width=0.45 \textwidth]{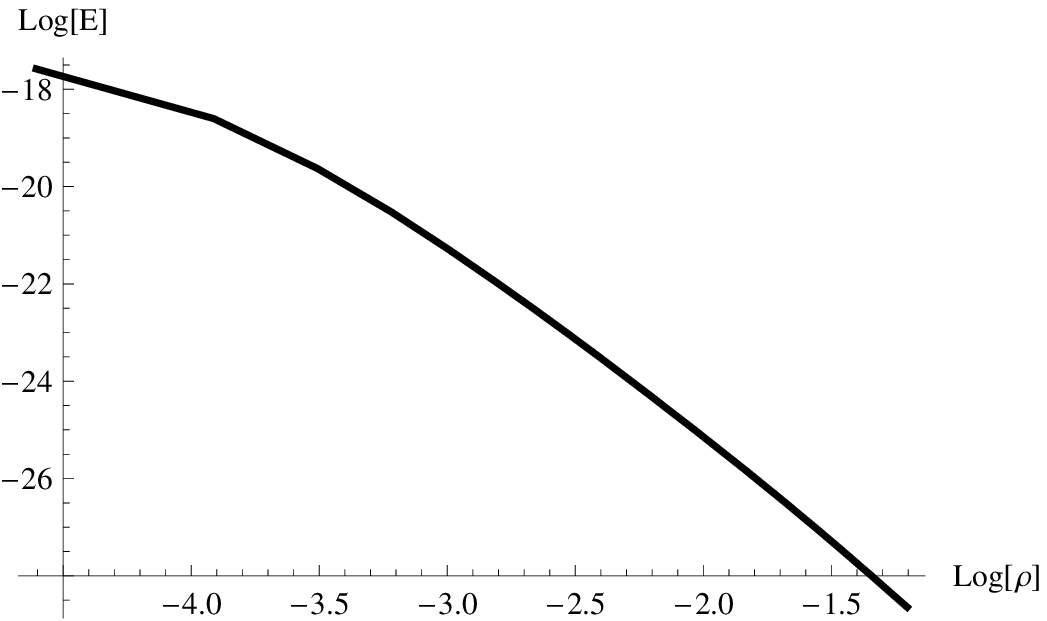}\\
\hskip 5cm (a)& \hskip 3cm (b)
\end{tabular}
\caption{(a) The noise energy for the SQUID plotted against various asymmetry parameters. The solid line is for  $\rho = \kappa = \chi =0$, the dotted line is for $\rho = 0.2, \kappa = \chi =0$, the dashed line is for $\kappa = 0.5, \rho = \chi =0$ and the dot-dashed line is for $\kappa = 0.5, \rho = 0.2$ and $\chi = 0$. This tells us that putting in some resistance asymmetry not only minimizes the noise energy globally but the noise at the resonance positions is also suppressed. (b) Plot of noise energy vs. resistance asymmetry. This is a monotonically decreasing curve for $\rho<0.3$ which tells us that noise energy can made very small by making the junction resistances a bit asymmetric.}
\end{figure}

\section{Discussion}
In this paper, we have studied the effect of asymmetry on the noise characteristics in high T$_C$ dc SQUIDs which behave as transmission lines.
It was shown that asymmetry can be tuned to improve the signal to noise ratio. In particular, some resistance asymmetry can cause a marked decrease in noise energy both globally as well as at the resonance positions and hence, it is not necessary to strive for extreme symmetry between the junctions during device design.

\section*{Acknowledgments}
\small{US would like to acknowledge Ed Tarte for useful initial discussions. FKW acknowledges support from NSERC through a discovery grant.  Research at Perimeter Institute is supported by the Government of Canada through Industry Canada and by the Province of Ontario through the Ministry of Research \& Innovation. US and AS gratefully acknowledge IISc, Bangalore, India for hospitality during the last stages of this project. }

\begin{appendix}
\section{Calculation details}
Adding and subtracting equations (\ref{normalised1}) and (\ref{normalised2}) and denoting $\theta_1 + \theta_2$ by $S$ and $\theta_1 - \theta_2$ by $D$, we get:
\begin{eqnarray}\label{sum}
\beta_C S'' + \chi \beta_C D'' &=&  i_B - S^{'}-\rho D^{'} - 2 \sin (S/2) \cos (D/2)\nonumber \\ && - 2 \kappa \sin (D/2) \cos (S/2) + \delta i_B
\end{eqnarray}
and

\begin{eqnarray}\label{diff}
\beta_C D'' + \chi \beta_C S'' &=&  j - D^{'} (1 + 2 \gamma) -\rho S^{'}- 2 \sin (D/2) \cos (S/2) \nonumber \\ && ~~~~~~~~~~~~ - 2 \kappa \sin (S/2) \cos (D/2) + \frac{4}{\beta} \delta \phi
\end{eqnarray}
Noise currents $i_{n1}$ and $i_{n2}$ have been combined to form the current noise $\delta i_B = i_{n1} + i_{n2}$ and flux noise $\delta \phi = \frac{\beta}{4} (i_{n1} - i_{n2}).$ \cite{Enpukunoise}

Let $S_0$ and $D_0$ satisfy equations (\ref{sum}) and (\ref{diff}) with $i_{n1} = i_{n2} = 0$ i.e. in the absence of noise. In the following steps, we obtain an expression for a low-frequency component of the voltage $v_S$ in the presence of noise where $v_S$ is defined by $ v_S = \frac{1}{2} \frac{dS}{dt}$. We consider the case where the noise current $\delta i_B$ and the noise flux $\delta \phi$ are small and express $S$ and $D$ as:

\be\label{pert1}
S = S_0 + S_n
\ee

\be\label{pert2}
D = D_0 + D_n
\ee
where $S_n$ and $D_n$ represent variations due to the noises $\delta i_B$ and $\delta \phi$.

Substituting equations (\ref{pert1}) and (\ref{pert2}) into equations (\ref{sum}) and (\ref{diff}), we obtain the following linearized equations for $S_n$ and $D_n$:

\begin{eqnarray}\label{linear1}
(S_n +\kappa D_n) \cos (\frac{S_0}{2}) \cos (\frac{D_0}{2}) &+& (D_n +\kappa S_n) \sin (\frac{S_0}{2}) \sin (\frac{D_0}{2})\nonumber \\
&=& -\beta_C S_{n}^{''} - \chi \beta_C D_{n}^{''} + \delta i_B - S_{n}^{'} - \rho D_{n}^{'}
\end{eqnarray}

\begin{eqnarray}\label{linear2}
(S_n - \kappa D_n) \sin (\frac{S_0}{2}) \sin (\frac{D_0}{2}) &-& (D_n +\kappa S_n) \cos (\frac{S_0}{2}) \cos (\frac{D_0}{2})\nonumber \\
&=& \beta_C D_{n}^{''} + \chi \beta_C S_{n}^{''} + (1 + 2 \gamma) D_{n}^{'} +  \rho S_{n}^{'} - \frac{4}{\beta} \delta \phi\nonumber \\
\end{eqnarray}
with

\be\label{trig1}
\cos (\frac{S_0}{2}) \cos (\frac{D_0}{2}) = \displaystyle\sum_{m} A_m \exp (imv_{dc}t )
\ee

\be\label{trig2}
\sin (\frac{S_0}{2}) \sin (\frac{D_0}{2}) = \displaystyle\sum_{m} B_m \exp (imv_{dc}t )
\ee
where it is assumed that $S_n \ll 1$ and $D_n \ll 1$. The trigonometric products in equations (\ref{trig1}) and (\ref{trig2}) have been expressed in a Fourier transform since the Josephson current has frequency components of $\omega = m v_{dc}$ in a finite voltage state of $v_{dc}$.  Here, $A_m$ and $B_m$ are coefficients representing the magnitude of the $m^{th}$ harmonics and m is an integer. Since equations (\ref{linear1}) and (\ref{linear2}) are linear, one can consider independently solutions of $S_n$ and $D_n$ for individual frequency components of of $\delta i_B$ and $\delta \phi$. Fourier transforms of $\delta i_B$ and $\delta \phi$ are defined as $\delta I_B (\omega)$ and $\delta \Phi (\omega)$ respectively. The voltage noise is $ \delta v_S =\displaystyle {1\over 2}\frac{dS_n}{dt}$ and its Fourier transform is $\delta V_S (\omega)$. Therefore, the voltage noise power spectrum is given by \cite{Enpukunoise} :

\be\label{noisepower}
S_V (\Omega) = \bigg < \delta V_S (\Omega) \delta V_{S}^{*} (\Omega) \bigg >
\ee
Next, we obtain the low-frequency component of the voltage noise, i.e. $\delta V_S (\Omega)$ where the frequency $\omega = \Omega$ is considered to be much lower then the Josephson oscillation frequency $\omega = v_{dc}$. It can be shown from equations \ref {linear1} - \ref{trig2} that the low frequency voltage noise $\delta V_{S,L} (\Omega)$ arises not only from low-frequency components of $\delta I_{B} (\Omega)$ and $\delta \Phi (\Omega)$, but also from high frequency components of $\delta I_{B} (\Omega - mv_{dc})$ and $\delta \Phi (\Omega - mv_{dc})$. The voltage noise due to high frequency components of $\delta I_{B} (\Omega - mv_{dc})$ and $\delta \Phi (\Omega - mv_{dc})$ has been expressed as the noise due to the Josephson mixing effect.

First, we obtain the value of $\delta V_{S} (\Omega)$ due to the low frequency components of $\delta I_{B} (\Omega)$ and $\delta \Phi (\Omega)$. It is difficult to solve equations \ref{linear1} and \ref{linear2} exactly for frequency components of $\delta I_B (\Omega)$ and $\delta \Phi (\Omega)$. However, since the frequency $\omega = \Omega$ is much lower than the Josephson oscillation frequency $\omega = v_{dc}$, one can regard $\delta I_B (\Omega)$ and $\delta \Phi (\Omega)$ as quasi-static changes of $I_B$ and $\Phi$ respectively. In this case, the value of $\delta V_S (\Omega)$ should be given by the change of the dc voltage $v_{dc }$ due to $\delta I_B (\Omega)$ and $\delta \Phi (\Omega)$, i.e.
\be\label{noiselowfreq}
\delta V_{S,L} (\Omega) = r_d \delta I_B (\Omega) + v_{\phi} \delta \Phi (\Omega)
\ee
where $r_d = \displaystyle\frac{dv_{dc}}{d i_b}$ and $v_{\phi} =\displaystyle \frac{dv_{dc}}{d\phi}$ are the dynamic resistance and the transfer function in the absence of noise respectively.

Next, we obtain the value of $\delta V_S (\Omega)$ due to high frequency components of $\delta I_B$ and $\delta \Phi$ i.e. $\delta V_{S,H} (\Omega)$ with $\omega = \Omega - mv_{dc}$. In this regime, the R.H.S of equations (\ref{linear1}) and (\ref{linear2}) are much larger than the L.H.S. Thus as a first approximation to obtain the lowest order perturbation solution, the L.H.S are set to zero. Taking Fourier transforms of equations (\ref{linear1}) and (\ref{linear2}) then gives:

\be\label {sn0}
\tilde S_{n}^{0} = {1 \over P} \bigg [ {4 \over \beta} \delta \Phi (\omega) s_1 - \delta I_{B} (\omega) d_1 \bigg ]
\ee

and

\be\label {dn0}
\tilde D_{n}^{0} = -{1 \over P} \bigg [ {4 \over \beta} s_2 \delta \Phi (\omega) - \delta I_B (\omega) s_1 \bigg ]
\ee
 where,
 \be
 P = s_{1}^{2} - d_{1} s_{2} \nonumber
 \ee
 with

  \be
  s_1 = i \omega \rho - \omega^2 \chi \beta_C \nonumber
 \ee

\be
 d_1 = i \omega ( 1+ 2\gamma) - \omega^2 \beta_C \nonumber
 \ee
  and
  \be
  s_2 = i \omega - \omega^2 \beta_C \nonumber
  \ee

Equations (\ref{sn0}) and (\ref{dn0}) are the zeroth order solutions. In order to get the first order solutions, we plug (\ref{sn0}) and (\ref{dn0}) into (\ref{linear1}) and (\ref{linear2}) L.H.S which gives expressions for $ \tilde S_{n}^{1}$ and $\tilde D_{n}^{1}$ from (\ref{linear1}), (\ref{linear2}), (\ref{trig1}), (\ref{trig2}), (\ref{sn0}) and (\ref{dn0}). Now, $\displaystyle \delta v_S = {dS_n \over 2dt} \Rightarrow \delta V_S (\omega) =\displaystyle {- i \omega \tilde S_{n} (\omega) \over 2}$. Therefore,

\be\label{noisehighfreq}
\delta V_{S,H} (\omega) = -{i \omega \over 2} \left[\tilde S_{n}^{0} (\omega) + \tilde S_{n}^{1} (\omega) \right]
\ee

The expression for high frequency noise with the expressions for $S_{n}^{0}$ and $S_{n}^{1}$ substituted in (\ref{noisehighfreq}) is quite long so we avoid writing the complete expression here. From (\ref{noiselowfreq}) and (\ref{noisehighfreq}) we get at the measurement frequency $\Omega$,

\be\label{noise}
\delta V_{S} (\omega = \Omega) = \delta V_{S,H} (\omega = \Omega) + \delta V_{S,L} (\omega = \Omega)
\ee

From (\ref{noisepower}), the average expectation value of (\ref{noise}) thus gives the expression for voltage noise power spectrum. In (\ref{trig1}) and (\ref{trig2}), we have taken $A_1 = A_{-1} = \cos ({\pi \phi \over 2})$ and $B_{1} = -B_{-1} = -i \sin ({\pi \phi \over 2})$ \cite{Enpukunoise}.

\end{appendix}


\begin{thebibliography}{99}
\bibitem{notebook} The mathematica notebook for the noise calculation can be provided on request.
\end{thebibliography}
\end{document}